# Temperature dependence of transport mechanisms in organic multiferroic tunnel junctions

Can Xiao[1], Huawei Sun[1], Luming Cheng[1], Xavier Devaux[2], Anthony Ferri[3], Weichuan Huang[4], Rachel Desfeux[3], Xiao-Guang Li[4], Sylvie Migot[2], Mairbek Chshiev[5], Sajid Rauf[1], Yajun Qi[1], Ruilong Wang[1], Tianjin Zhang[1], Changping Yang[1], Shiheng Liang[1,2*] and Yuan Lu[2*]

[1] *Faculty of Physics and Electronic Science, Hubei University, Wuhan 430062, P. R. China*
[2] *Institut Jean Lamour, UMR 7198, CNRS-Université de Lorraine, Campus ARTEM, 2 Allée André Guinier, BP 50840, 54011 Nancy, France*
[3] *Univ. Artois, CNRS, Centrale Lille, ENSCL, Univ. Lille, UMR 8181, Unité de Catalyse et Chimie du Solide (UCCS), F-62300 Lens, France*
[4] *Hefei National Laboratory for Physical Sciences at Microscale, Department of Physics, University of Science and Technology of China, Hefei 230026, P. R. China*
[5] *Univ. Grenoble Alpes, CEA, CNRS, Grenoble INP, INAC-Spintec, 38000 Grenoble, France*

**\*Corresponding authors:**
E-mail: yuan.lu@univ-lorraine.fr; shihengliang@hubu.edu.cn

**Abstract:**

Organic multiferroic tunnel junctions (OMFTJs) with multi-resistance states have been proposed and drawn intensive interests due to their potential applications, for examples of memristor and spintronics based synapse devices. The ferroelectric control of spin-polarization at ferromagnet (FM)/ferroelectric organic (FE-Org) interface by electrically switching the ferroelectric polarization of the FE-Org has been recently realized. However, there is still a lack of understanding of the transport properties in OMFTJs, especially the interplay between the ferroelectric domain structure in the organic barrier and the spin-polarized electron tunneling through the barrier. Here, we report on a systematic study of the temperature dependent transport behavior in $La_{0.6}Sr_{0.4}MnO_3$/PVDF/Co OMFTJs. It is found that the thermal fluctuation of the ferroelectric domains plays an important role on the transport properties. When T>120K, the opposite temperature dependence of resistance for in up and down ferroelectric polarization states results in a rapid diminishing of tunneling electroresistance (TER). These results contribute to the understanding of the transport properties for designing high performance OMFTJs for memristor and spintronics applications.

# 1. Introduction

The organic multiferroic tunnel junctions (OMFTJs) [1,2,3,4] have recently attracted much attention due to their synergetic advantages of spintronics, organic electronics, and ferroelectric electronics. In comparison to inorganic based devices, organic materials are appealing because of the long spin lifetime of charge carriers [5], low cost and chemical diversity [6,7]. In an OMFTJ, the core structure comprises a ferroelectric organic (FE-Org) tunnel barrier layer sandwiched between two ferromagnetic (FM) electrodes. Owing to the combined tunneling magnetoresistance (TMR) and tunneling electro-resistance (TER) effects [8,9,10], it shows a performance of a multi-resistance switching behavior. Therefore the magnetic and ferroelectric dependent nonvolatile multi-states properties can be utilized to develop spintronics-based artificial intelligence devices, such as synapse [10,11]. More interestingly, it has been demonstrated that the spin-polarization at the FM/FE-Org interface, such as "spinterface" can be inverted by electrical switching of the ferroelectric polarization of the FE-Org [1]. This discovery emphasizes the critical role of "spinterface" [12,13] on the determination of the spin polarization at the organic/ferromagnetic interface and opens up new functionality in controlling the injection of spin polarization into organic materials via modulating the ferroelectric polarization of the barrier.

In the inorganic all-perovskite multiferroic tunnel junction (MFTJ), the temperature dependence of transport properties and mechanisms have been studied [14,15]. The structural defects such as oxygen vacancies in the oxide ferroelectric barriers were found to lead to a thermally activated inelastic hopping through chains of localized states in thicker barriers and at higher temperatures. This conduction channel is not sensitive to the ferroelectric orientation and cannot carry spin polarization information, which reduces the TMR and TER effects significantly at higher temperatures [16]. For OMFTJ, the replacement of inorganic ferroelectric barrier by organic polymer material should introduce different mechanisms to influence the temperature dependent transport behavior. However, up to now, there is still a lack of illumination on this issue. In this work, we have measured temperature dependence of transport behavior in OMFTJs, which shows different mechanisms dominate the transport properties at different temperature ranges. The understanding of the transport properties is essential for designing advanced OMFTJs for memristor and spintronics applications.

**2. Experimental methods**

In our experiments, OMFTJ samples based on $La_{0.6}Sr_{0.4}MnO_3$ (LSMO)/Polyvinylidene fluoride (PVDF)/Co/Au structure were fabricated. The sample structure is shown in **Figure 1(a)**. LSMO film with thickness of ~85 nm was firstly epitaxial grown on <100> oriented $SrTiO_3$ (STO) substrates at 750°C using DC magnetron sputtering. The film was subsequently annealed at 800°C for two hours in $O_2$ atmosphere before being slowly cooled down to the room temperature. For the devices prepared for transport measurement, the LSMO layer was etched by using hydrogen chloride (37%) to pattern 200μm width bars as the bottom electrodes. The PVDF film was prepared by spin-coating method. The solution was obtained by dissolving PVDF powders (purchased from Sigma-Aldrich) into N,N-dimethylformamide (DMF) (purchased from Sigma-Aldrich). The solution was then spin-coated onto the LSMO/STO(001) substrate with a speed of 3000 RPM for 1min. Subsequently, the as-coated film was annealed at 150°C in air for two hours to improve the crystallinity of the ferroelectric *β*-phase. The thickness of the PVDF film was controlled by adjusting the concentration of the DMF:PVDF solution. With the concentration of 20mg/mL, the thickness of PVDF is estimated to be about 20 nm. After spin-coating of PVDF layer, the 10nm Co/10nm Au was deposited by molecular beam epitaxy (MBE) with a shadow mask to form the top electrode. In the purpose of minimizing the metal diffusion into organic material, the temperature during the growth of the top electrode was maintained at ~80K by cooling the substrate with liquid nitrogen. The final junction, that is schematically shown in **Figure 1(a)**, has a typical size of about 200×200 μm$^2$.

HR-STEM was performed by using a probe-corrected microscope JEOL ARM200F (cold FEG) equipped with a GATAN GIF Quantum energy filter to reveal the structure and element distribution in the FM/FE-Org interface. Thin lamella were prepared by conventional focused ion beam milling with FEI Helios Nanolab dual beam microscope. Pt-C deposit was used as capping layer. Due to the strong sensitivity of polymers to electron-beam damages, the microscope was operated at 80kV and the temperature of the sample was held at 103K using a Gatan double-tilt liquid nitrogen cooling sample holder. Electron energy loss spectroscopy spectrum images (EELS-SI) were denoised using a principal component analysis before to be processed. Before observation the lamella were plasma-cleaned under Ar-O$_2$ atmosphere during 2 minutes.

Piezoresponse force microscopy (PFM) was used to check the ferroelectric properties of the PVDF film. For this study, the PVDF film was spin-coated on a thermal oxidized SiO$_2$/*n*-Si substrate covered with 100nm Au. The surface morphology was studied using an atomic force microscopy (AFM) (Asylum Research/Oxford Instruments, MFP-3D, USA) working in contact mode under environmental conditions. Local electrical experiments were performed by using dual AC resonance tracking (DART) PFM.

The magneto-transport measurement was performed in a closed cycle cryostat by varying temperature from 10K to 300K in the presence of magnetic field up to 4 kOe. *I-V* measurements were performed using Keithley-2400 as a voltage source and a Keithley-6487 picoammeter to measure the current. Since junction resistance is much larger than electrode resistance, two terminal configuration has been used for *I-V* characterization. To polarize PVDF barrier, different amplitude of voltage pulses with a ramp of 0.1V/s and a duration of 1s was applied to the junction. It was verified with conductive AFM that the polarizing voltage around ±3V range does not induce damages to organic barrier with the formation of pinholes inducing leakage currents.

## 3. Results and discussion

The surface morphology of PVDF barrier layer has been investigated by AFM observation. It reveals a smooth surface with root mean square (RMS) roughness <3.5 nm in 1×1 μm$^2$ scanning range, as shown in **Figure 1(b)**. The surface morphology exhibits a feature of grains composed by needle-like crystallites with a width of ~30 nm and length of ~100 nm, which is characteristic of the ferroelectric *β*-phase of PVDF [17]. The ferroelectric properties of the PVDF barrier have been characterized by PFM to check the homogeneity in ferroelectric polarization switching. The PFM phase map was recorded over 10×10 μm$^2$ PVDF surface area, as shown in **Figure 1(c)**. The film was initially polarized with a biased tip at positive voltage (+3V), then an opposite poling bias (-3V) over an inside area of 5×5 μm$^2$. As observed, the clear switching of the polarization is obtained by applying positive and negative bias voltages, suggesting a homogeneous polarization switching.

**Figure 2(a)** shows a bright-field STEM image of the sample. The bottom LSMO magnetic electrode revealed a good homogeneity with a thickness of ~85 nm and low roughness of about 3 nm. The spacer layer PVDF appears homogenous without pinholes with a thickness of ~20 nm. This also validates the homogeneous ferroelectric switching measured by PFM phase mapping (**Figure 1(c)**). In addition, we have performed the chemical analysis by using EELS to check the Co diffusion in the organic barrier. **Figure 2(b)** shows an high angle annular dark-field (HAADF) STEM image where EELS spectrum image was recorded. For the three regions marked with orange, red and blue squares on the STEM image, EELS spectra are collected and presented in **Figure 2(c)**. Although the Co signal is well distinguished in the region close to Co/PVDF interface (orange zone), we cannot find any Co signal in the middle of PVDF barrier (red zone). There are also no La and Mn signals in PVDF barrier compared to the strong signals from the region close to PVDF/LSMO interface (blue zone). The chemical EELS maps of Co/PVDF/LSMO interface is presented in **Figure 2(d)**. The maps of chemical elements Co (yellow), F (red) and La (sky-blue) are extracted from the EELS spectrum images by using signals of $Co_{L3}$(779eV), $F_K$(685eV) and $La_{M5}$(820eV) edges, respectively. It also appears that there is very small Co diffusion inside the organic barrier and the interface PVDF/LSMO is quite sharp. All these results prove that there is very limited Co diffusion into the PVDF organic barrier when Au/Co electrode is deposited at low temperature, which agrees well with our previous report in P(VDF-TrFE) based OMFTJ [4].

**Figure 3(a)** shows the *I-V* characteristic of the device measured at different temperature under a magnetic field of 2kOe (in parallel state). The device has been only initially polarized by +1.5V at 15K. It can be seen that the *I-V* curves show non-linear properties which indicates a tunneling behavior. We have also plotted the bias dependent differential resistance (d*V*/d*I*) as a function of temperature in **Figure 3(b)**. It is found the resistance increases at small bias, which can be understood due to the deformation of barrier height at larger bias. It is interesting to find that the temperature dependent behavior is different for small and large bias. **Figure 3(c)** displays the differential resistance as a function of temperature for zero bias and +0.5V bias. At zero bias, the resistance varies in three temperature regions: firstly decreases from 15K to 40K, and then increases with the increase of temperature to 120K following by a decrease up to 300K. However, the resistance at +0.5V bias shows a monotonous increase with the temperature. This could be due to the influence of large bias on the ferroelectric domain in PVDF barriers, and this effect could be more pronounced at high temperature [18]. In the following part, we will discuss more details for the temperature dependent resistance.

**Figure 4** shows the magneto-response loops measured at 10K after different polarizations (+1.2V for positive polarization and -1.5V for negative polarization, respectively). When scanning the magnetic field, the magnetization of LSMO and Co are switched separately to get a parallel or antiparallel magnetization configurations and results in the tunneling magnetoresistance curves. The *TMR* is defined by the following relations: $TMR = \frac{R_{AP}-R_P}{R_P} \times 100\%$, where $R_P$ and $R_{AP}$ are junction resistance where the magnetizations of two electrodes are parallel and antiparallel, respectively. It is clearly that the junction shows a positive *TMR* of +8.3% when the PVDF is positively polarized and a negative *TMR* of -15.3% when the PVDF is negatively polarized. At the same time, a clear difference between the two parallel resistances ($R_P$) for the two polarizations results in a *TER* about 65%. The *TER* is defined as: $TER = \frac{R_P^{Down}-R_P^{Up}}{R_P^{Up}} \times 100\%$, where $R_P^{Down}$ and $R_P^{Up}$ are parallel resistance in PVDF "down" and "up" polarization states, respectively. The four resistance states associated with different magnetization and ferroelectric configurations are identified in the inset of **Figure 4**, which clearly demonstrates the MFTJ function of our LSMO/PVDF/Co organic tunnel junction. To validate that the measured TMR is not due to the anisotropic magnetoresistance (AMR) of FM electrodes, we have checked the AMR of LSMO and Co electrodes, which is much smaller than the observed TMR (see **Supplementary S1**).

The change of TMR sign with different PVDF polarizations can be understood as due to the sign change of the spin-polarization at the organic/ferromagnetic "spinterface" depending on the ferroelectric polarization of the organic barrier [1]. Due to the strong hybridization at organic/ferromagnetic interface, there exist strong spin-dependent metal induced gap states (MIGS) inside PVDF at Co/PVDF interface. The decay rate of spin up and spin down MIGS can be changed upon the ferroelectric polarization states, which results in a sign change in spin-polarization in different PVDF layer. As shown in the schematics in **Figure 1(d) and 1(e)**, the sign of the spin-polarization of the second C layer can be changed due to the switching of PVDF polarization, which acts as a final spin-filter for electrons tunneling through the PVDF barrier by considering Jullière's model [19]. We have verified the stability of ferroelectric repetition and the reproductivity of TMR sign change with another device with a thinner PVDF thickness (15mg/mL) (see **Supplementary S2**).

**Figure 5(a)** shows the magneto-response loops at different temperature for both polarization states. We have firstly polarized PVDF in both polarization states at 10K and then increased the temperature separately in each polarization state. It is found that the TMR values in both polarization states decrease monotonously when increasing the temperature (**Figure 5(b)**). The positive TMR completely vanishes at temperature higher than 150K while the negative TMR can persist even at 250K. This behavior generally occurs for the organic tunneling junction using the LSMO electrode [1,13,20]. The reduction of MR at a high temperature can be understood by the reduction of LSMO interface spin-polarization in addition with the decrease of PVDF spin diffusion length [4]. **Figure 5(c)** shows the variation of TER with temperature. It is found that the TER reveals a much complicated temperature dependent behavior. It firstly increases from 65% at 10K to a maximum of 80% at 100K, then following with a rapid decrease to 5% up to 300K.

To understand the temperature dependent TER and transport mechanisms, we have studied the temperature dependent resistances in both polarization states, as shown in **Figure 6(a)**. To avoid the influence of large bias on the ferroelectricity of PVDF, all resistances are measured with a small bias of +10mV without performing *I-V* measurement. It can be observed that after positive polarization at 10K, the resistance firstly decreases slightly with temperature from 10K to 60K (green region) and then increases with the increase of temperature to 120K (yellow region), followed by a decrease up to 290K (grey region). This behavior is quite similar as the tendency of d$V$/d$I$ measured at zero bias shown in **Figure 3(c)**. For the case after negative polarization at 10K, the resistance firstly decreases to a minimum at 60K and then increases monotonously with temperature until to 290K. Below 120K, the resistance in both polarization states varies in the same tendency with the temperature and there is small variation in TER. However above 120K, the opposite change tendency of resistance in both polarizations induces a rapid decrease of TER for the OMFTJ device.

To explain this temperature dependence of transport behavior, we have defined three temperature ranges. For the first range from 10K to 60K (green region), the decrease of resistance with increasing temperature could be due to the activation of the hopping process due to the inelastic tunneling which reduces the effective tunneling resistance [21]. Since the thickness of PVDF organic barrier is about 20 nm, it is not possible for electrons to directly tunnel through the barrier. Hopping process through the defect states close to Fermi level inside the barrier is indispensable. Thermal activated phonon can help electron hopping through the defect sites and effective reduce the tunneling resistance [21,22]. Kondo effect due to the magnetic impurity scattering can also effectively increase the junction resistance at low temperature [23]. However, this mechanism can be excluded in our case because that this effect will also manifest zero-bias anomaly (ZBA) in the I-V curves at low temperature, but we do not observe this ZBA in our measurements shown in **Figure 3(b).**

From 60K to 120K (yellow region), the increase of resistance with temperature for both polarization states can be due to several mechanisms. First of all, the metallic behavior induced by pinholes inside the barrier can result in such behavior [24]. However, as observed in our TEM, the barrier is almost homogenous and no pinhole is observed. To definitively confirm this point, we have performed conductive AFM characterization. No conducting path is obtained for an applied bias of +5V, as demonstrated by the uniform contrast associated to insignificant background current (~30pA) as shown in **Supplementary Figure S3**. The second possibility is that PVDF has a semi-crystalline nature which is known for the ubiquitous presence of amorphous phase in addition to the ferroelectric crystalline phase [25]. When temperature increases, thermal energy can result in a structural disorder in PVDF film, as already demonstrated that in some organic superconductors the structure transition exists around 80K [26]. The thermal activation of structural disorder could induce an increase of electron scattering and result in the increase of resistance. However, this hypothesis can be also excluded because the structure disorder will also induce a disorder in ferromagnetic domains, which should manifest a decrease of resistance in the positive polarization state with the increase of temperature. The third possibility could be due to the thermal strain effect. **Supplementary Figure S4** displays the saturation magnetization ($M_s$) as a function of temperature for the PVDF/Co bilayers and pure Co. We have observed an increase of magnetization about 10% from 10K to 300K for the Co on PVDF. This behavior is completely different from the normal decrease of $M_S$ with temperature for the pure Co due to the magnetization disorder generated by the limited Curie temperature. In fact, the increase of Co $M_s$ is related to the magnetostriction effect of Co because PVDF has an anisotropic thermal expansion, which can generate a uniaxial strain on Co when changing the temperature. The thermal strain can then effectively influence on the junction resistance [27], as discussed in **Supplementary S4**.

From 120K to room temperature (grey region), the temperature dependence can be explained by the thermal fluctuation of ferroelectric domains, as schematic shown in **Figure 6(b)**. In fact, every needle-like PVDF crystallite can act as an individual ferroelectric domain [4]. When the thermal energy is high enough to overcome the energy barrier ($E_b$) between the two polarization states [28], the chain of molecular starts to rotate to another polarization state. With the increase of temperature, more or more domains with opposite polarization nucleate in both polarization states, which results in a reduce of resistance for the positive polarization state while an increase of resistance for the negative polarization state. Since negligible electric field was applied during the increase the temperature, the evolution of ferrielectric domain of PVDF could be described by the Kolmogorov-Avrami-Ishibashi (KAI) model [18,29,30], which assumes the statistical formation of a large number of nucleation sites and homogeneous domain growth. The transition temperature ($T_{Tr}$) at 120K can allow us to roughly estimate the energy barrier between the two polarization states to be about $E_b=k_BT_{Tr}=10$meV. This value is one order smaller than the reported nucleation energy barrier for P(VDF-TrFE) thin film (0.4eV) grown by Langmuir-Blodgett method [31]. The smaller energy barrier could be related to the crystal inhomogeneities, lattice defects, interface defects or small residual nuclei [32,33,34]. With the increase of temperature, the opposite change tendency of resistance in both polarizations results in the disappearing of the contrast of the two polarization states, thus vanishing the TER effect. In addition, we have measured the temperature dependent magneto-transport behavior for another thin device of LSMO/PVDF(13nm)/Co (see **Supplementary S5**), which has been polarized by two different methods during the increase of temperature: 1) states only polarized at 20K and 2) states polarized at each measuring temperature. As shown in **Supplementary Figure S5**, the second polarizing method can indeed achieve better polarization states at higher temperature. The TER can be much enhanced at RT compared to the first polarizing method. This validates that the polarization relaxation by the thermal fluctuation dominates the transport properties in the range $T>120$K in the first polarizing method. In comparison, the influence of thermal activation of domain switching plays a negligible role in the temperature dependent behavior in inorganic MFTJ based on BaTiO$_3$ [14] due to the high nucleation barrier energy ($40k_BT$) [35].

## 4. Conclusions:

In summary, we have studied the temperature dependent transport behavior in La$_{0.6}$Sr$_{0.4}$MnO$_3$/PVDF/Co OMFTJs. We have found that different mechanisms can dominate the transport properties in different temperature ranges. Below 60K, the transport is dominated by electron hopping through the thick organic barrier, which shows a decrease of resistance with the increase of temperature in both polarization states. Between 60K and 120K, the thermal strain could induce an increase of resistance in both polarization states when increasing the temperature. When *T*>120K, it is found that the thermal fluctuation of the ferroelectric domains plays an important role on the transport properties. The opposite temperature dependence of resistance for both ferroelectric polarization states results in a rapid diminishing of tunneling electroresistance. However, polarizing the junction at high temperature can recover better polarization degree to enhance TER at room temperature. These results pave the ways to develop high performance OMFTJs for memristor and spintronics applications.

**Acknowledgements**

The authors thank Dr. A. Da Costa for technical support on AFM experiments. We thank also Prof. Zhi-Gang Yu for fruitful discussions. We acknowledge the support from the French National Research Agency (ANR) FEOrgSpin project (Grant No. ANR-18-CE24-0017-01) and SIZMO2D project (Grant No. ANR-19-CE24-0005-02). This work is supported by the National Natural Science Foundation of China (No. 11674086, 11904088), Natural Science Foundation of Hubei Province (No. 2019CFB183), Scientific Research Project of Education Department of Hubei Province (No. Q20191010) and Research Project of Wuhan Science and Technology Bureau (No. 2019010701011394). The "Région Hauts-de-France" and the "Fonds Européen de Développement Régional (FEDER)" under the "Contrat de Plan État-Région (CPER)" project "Chemistry and Materials for a Sustainable Growth" is acknowledged for funding of MFP-3D microscope. Experiments were performed using equipment from the platform TUBE–DAUM funded by FEDER (EU), ANR, the Region Lorraine and Grand Nancy.

**Figures:**

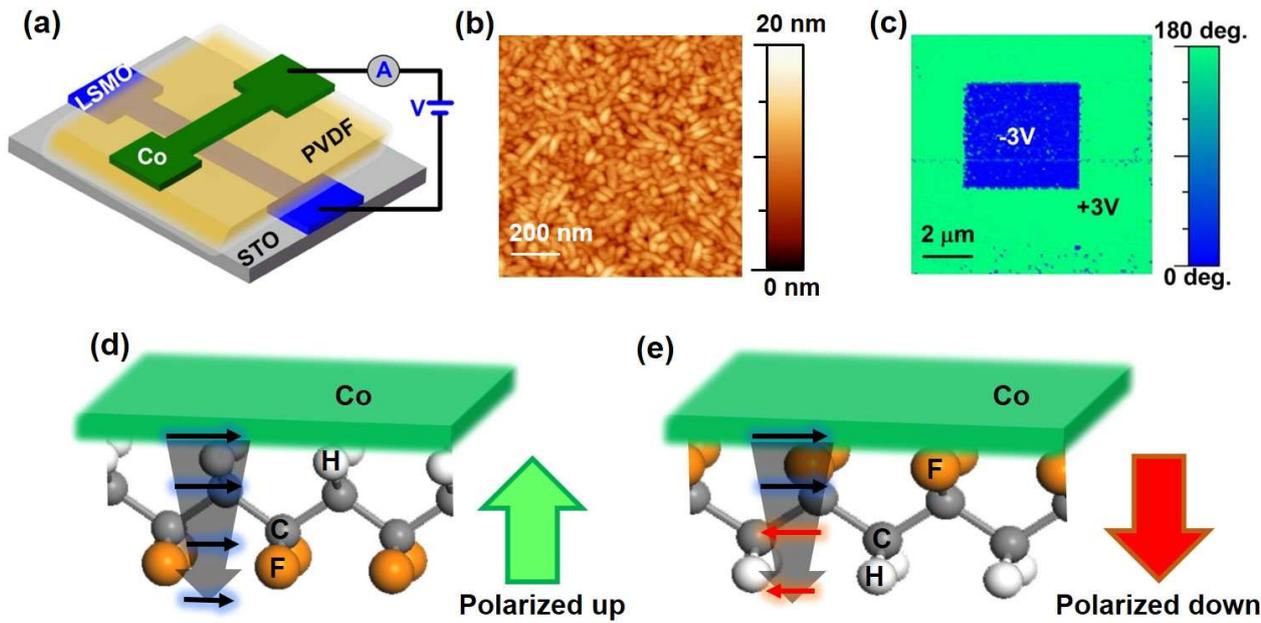

Figure 1: (a) Schematics of the LSMO/PVDF/Co device. (b) AFM topography measurement of the PVDF barrier surface over 1×1 μm² area. (c) PFM phase image recorded on the PVDF surface. Contrasts showing the ferroelectric switching are obtained after application of +3 V DC bias on the tip over 10×10 μm² and subsequently –3 V over inside 5×5 μm² area. (d-e) schematics of spin-filter for electron tunneling through PVDF barrier in polarized-up and polarized-down states.

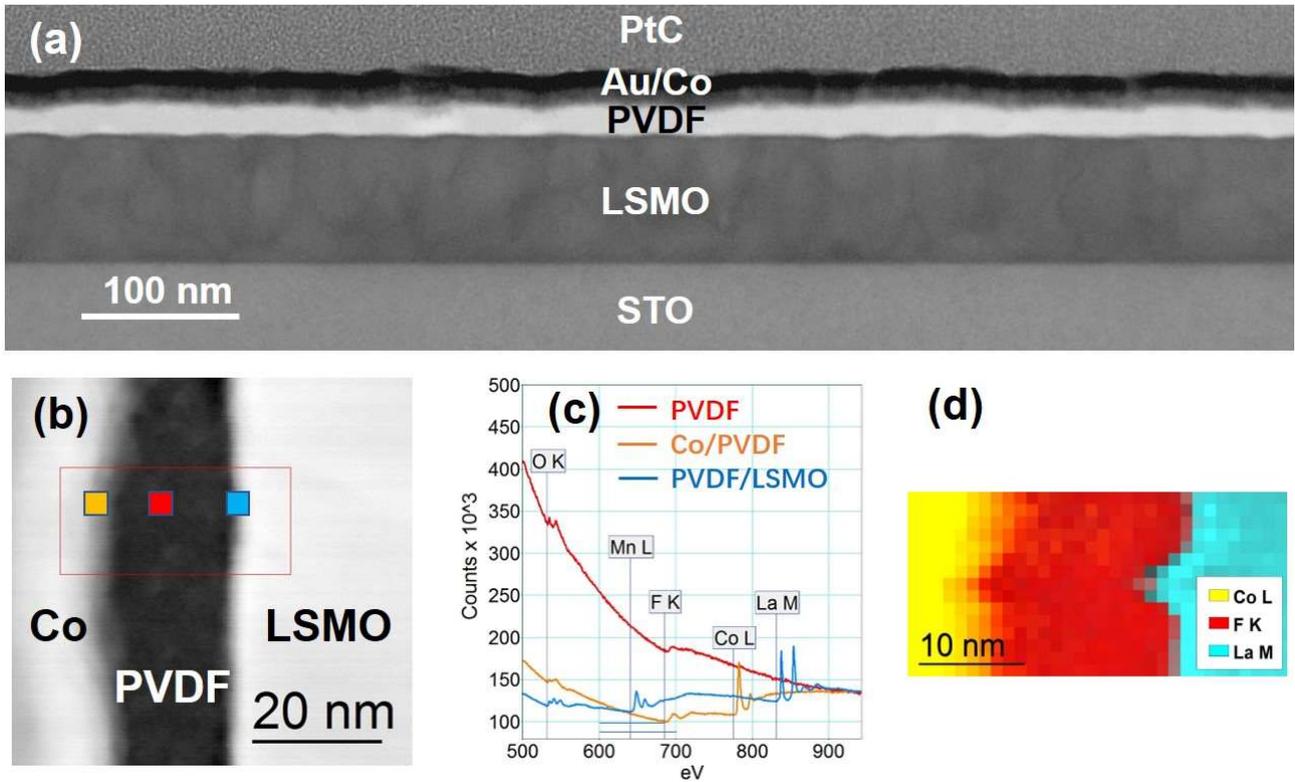

**Figure 2:** (a) Large scale bright-field STEM image of the LSMO/PVDF/Co device. The PtC layer deposited during the lamella preparation appears on the top of the image. (b) HAADF STEM image of the sample. The red frame indicates where an EELS-SI was recorded. The colored squares shows the area where EELS spectra in (c) were extracted. (c) Typical EELS spectra of the Co/PVDF interface, the PVDF layer and the PVDF/LSMO interface. (d) Elemental map drawn from quantitative analysis of EELS-SI.

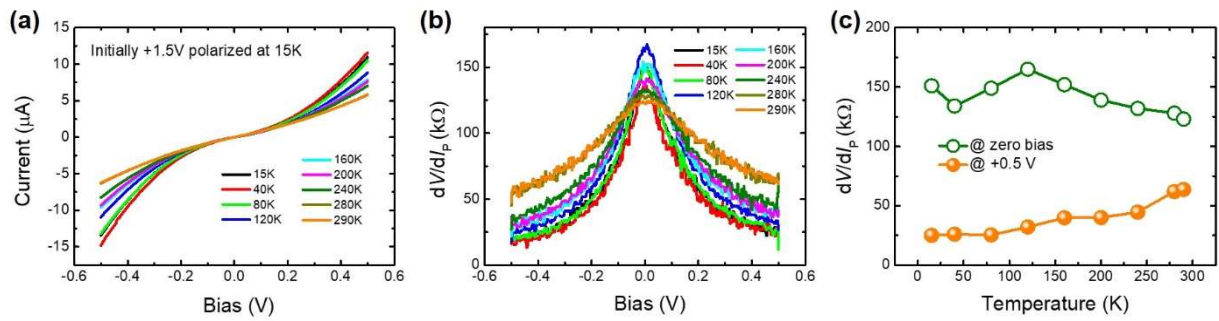

**Figure 3:** (a) Temperature dependence of *I-V* characteristics under the range between -0.5V to +0.5V. (b) Differential resistance as a function of bias at different temperatures. (c) Temperature dependence of differential resistance at zero bias and +0.5V.

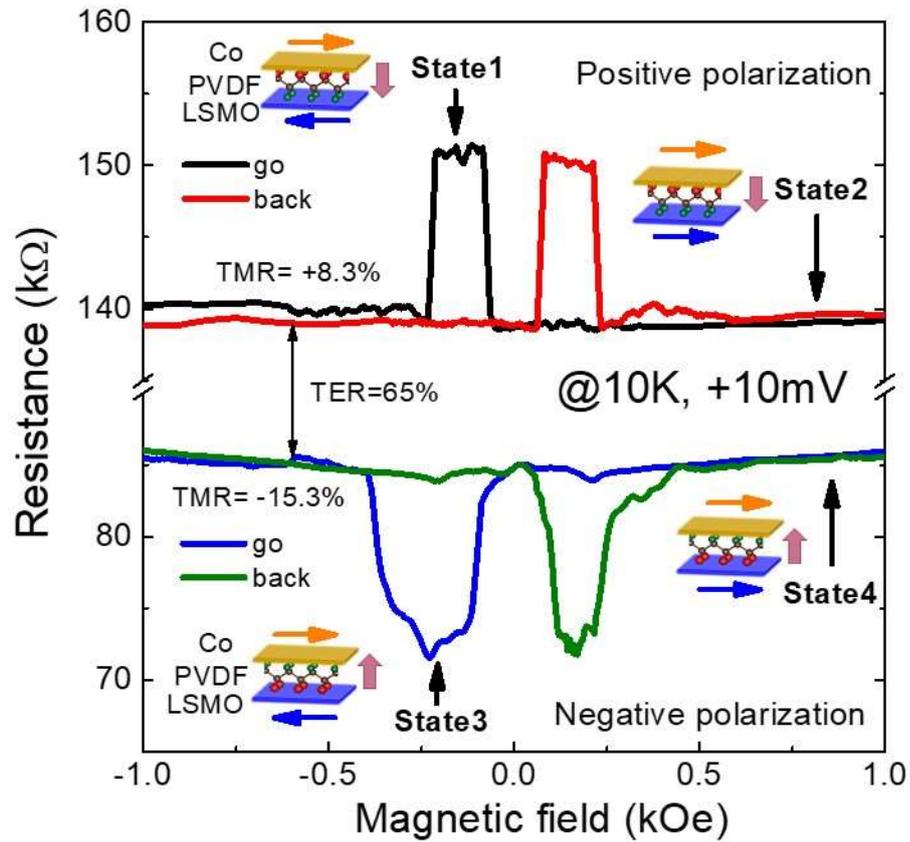

**Figure 4:** Magneto-response curves measured under a bias of +10 mV at 10 K after +1.2 V and −1.5 V polarizing. The insert schematics show four resistance states associated with different polarization and magnetization orientations.

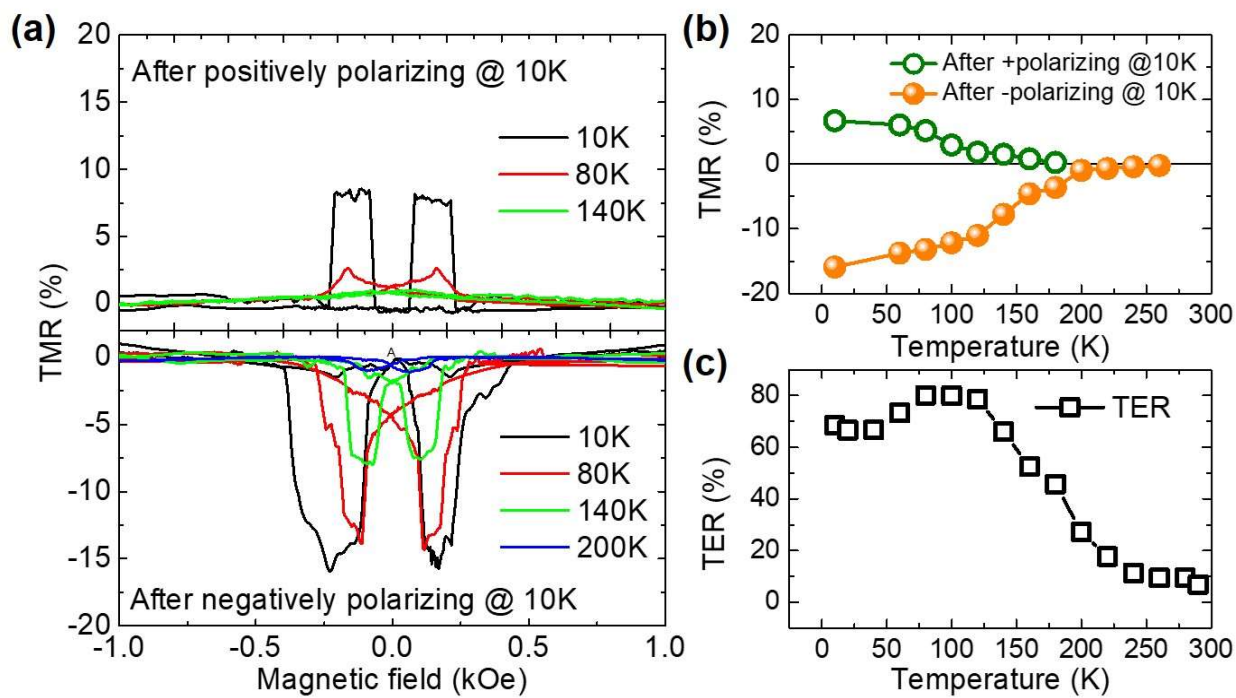

**Figure 5:** (a) Magneto-resistance loops measured at different temperatures for positive and negative polarization states. (b) Temperature dependence of MR in both polarization states. (c) Temperature dependence of TER.

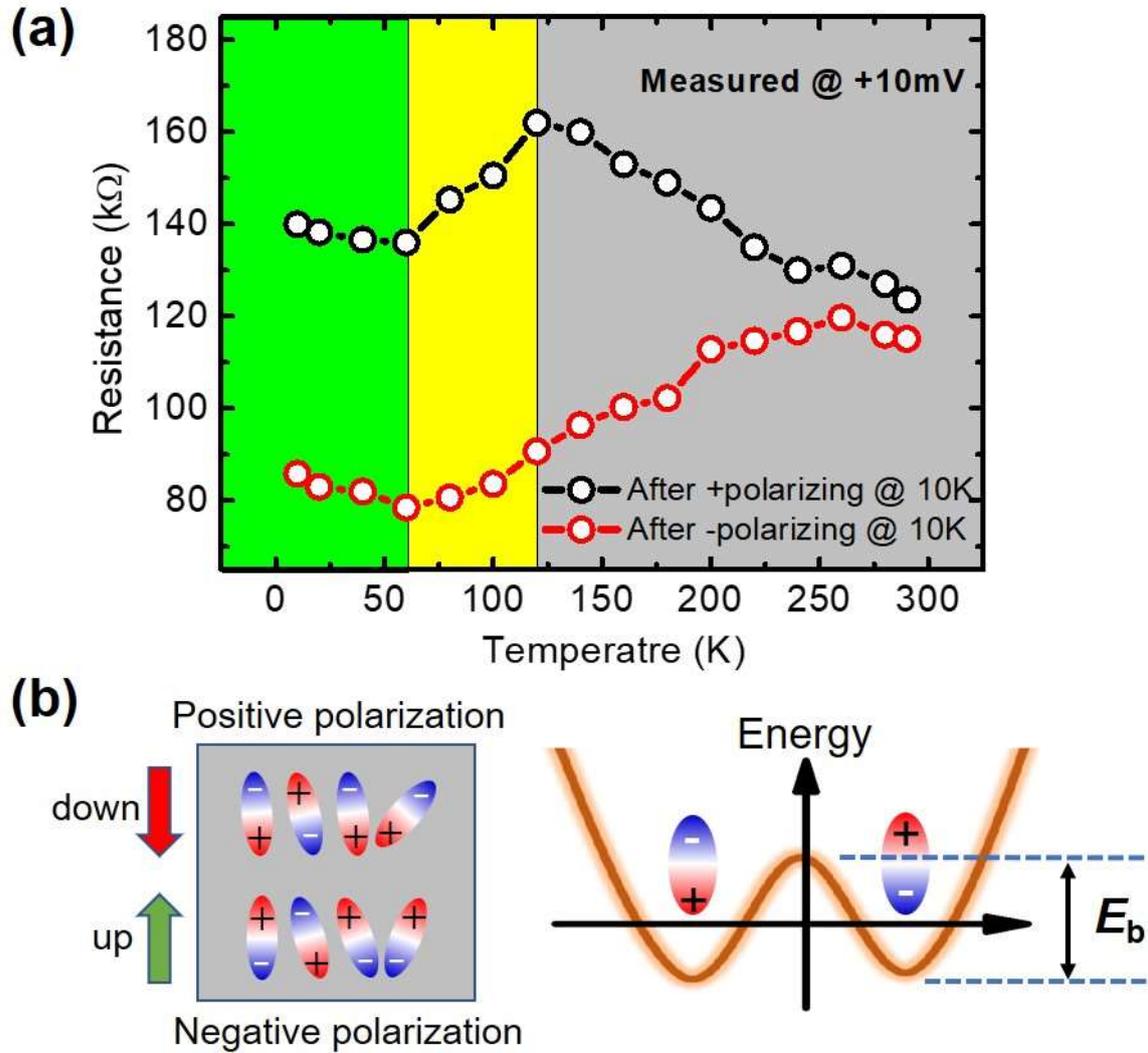

**Figure 6:** (a) Temperature dependence of $R_P$ in both polarization states. The resistance is measured with small bias of 10mV. (b) Schematics of polarized states in the PVDF barrier at temperature range between 120K and 300K (grey zone). The schematic of energy barrier between two ferroelectric polarization states under thermal fluctuation is also shown.

# Supplementary Information

**S1. Anisotropic magnetoresistance in the ferromagnetic electrodes**

To validate that the measured TMR is not due to the AMR of FM electrodes, we have checked the AMR of LSMO and Co electrodes. As shown in **Figure S1,** the AMR of LSMO and Co shows a much smaller value lower than 0.4%. However, the MR in our device can be as large as 15%, which is much larger than that of AMR.

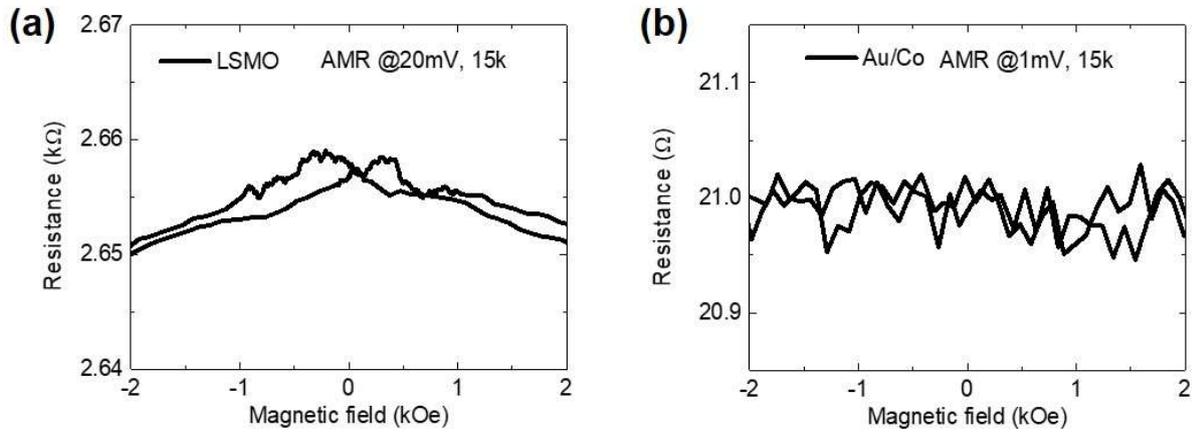

**Figure S1:** AMR measurement of (a) bottom LSMO and (b) top Au/Co electrodes.

**S2. Stability of ferroelectric repetition**

We have prepared another device with a thinner PVDF thickness (15mg/mL) to check the stability of ferroelectric switching. Compared to the thick PVDF sample (20mg/mL) presented in the main text, a smaller switching voltage of ±1V is enough to saturate the two polarization states due to the thinner thickness. As shown in **Figure S2(a) and S2(b)**, the TMR changes from +1.64% to -1.81% when the PVDF is switched from positive polarization state to negative polarization state. At the same time, a clear difference between the two parallel resistances for the two polarizations allows to reach a TER of about 14.3%. Although the amplitude of TMR and TER is smaller than that presented in the main text, the phenomena of ferroelectric modulation of spin-polarization can be reproduced in our OMFTJs with different PVDF thickness. Successive alternative voltages of ±1V are applied on the device to test the repetition stability. We have collected magnetoresistance curve after each polarizing pulse. As shown in **Figure S2(c) and S2(d)**, the parallel resistance and TMR can be robustly reproduced after each polarizing cycle.

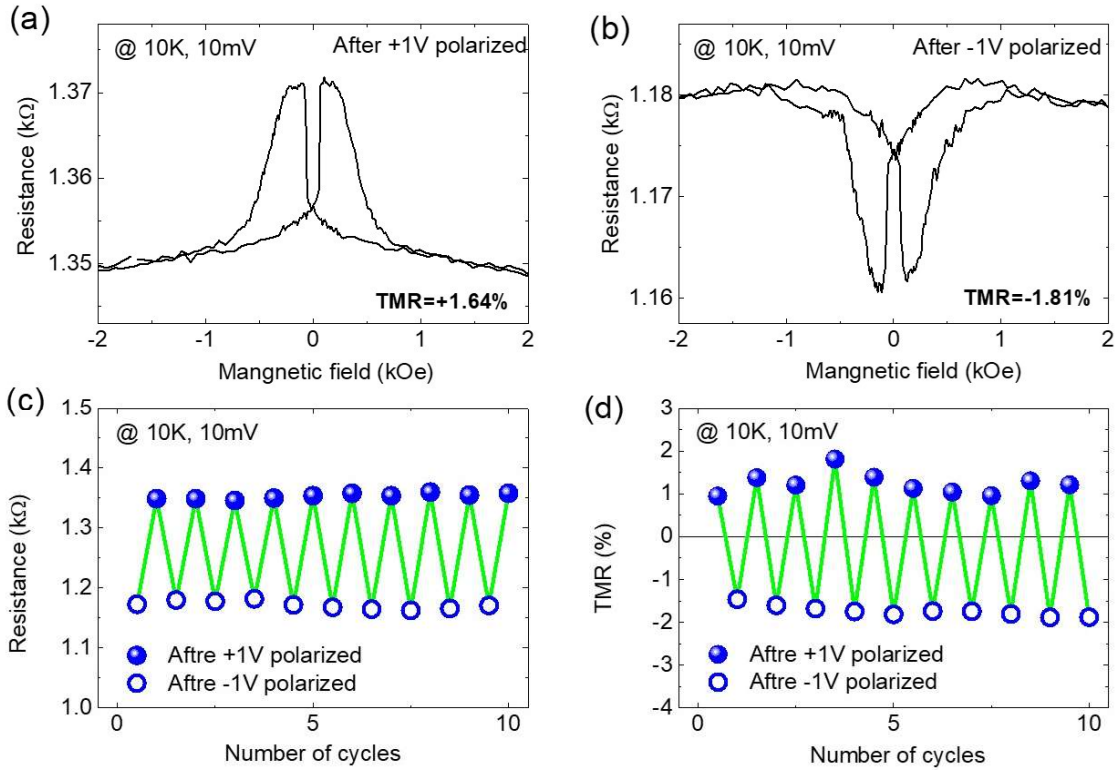

**Figure S2:** Tunneling magnetoresistance curves measured under 10mV at 10K after (a) +1V and (b) -1V polarizing of the 15mg/mL PVDF device. Repetition cycles after successive switching with ±1V voltage pulses for (c) parallel resistance and (d) TMR.

**S3. Possibility of pinhole**

Conductive AFM (c-AFM) measurement is the most direct way to identify the existence of pinholes. Nanoscale conductivity variations were probed through current mapping experiments using the c-AFM technique. A Ti/Ir-coated silicon tip and cantilever with a stiffness of 3N·m$^{-1}$ were used. DC bias voltages ranging from -10 and +10V between the grounded AFM tip and the bottom electrode were applied during the scan. The electrical conductivity was locally probed over the surface of the PVDF layer. The current map recorded over 20×20 μm² large area is displayed in **Figure S3** for samples with PVDF thicknesses of 20nm. No conducting path was obtained (here for an applied bias of +5V), as demonstrated by the uniform contrast associated to insignificant current (~30pA). Such absence of conduction signal was confirmed in all places measured regardless the applied bias voltage when varying between -10 and +10V. It confirms that the contribution of pinhole on transport is very limited. This also proves that the polarizing voltage in the ±3V range does not induce damages to the organic barrier with the formation of pinholes.

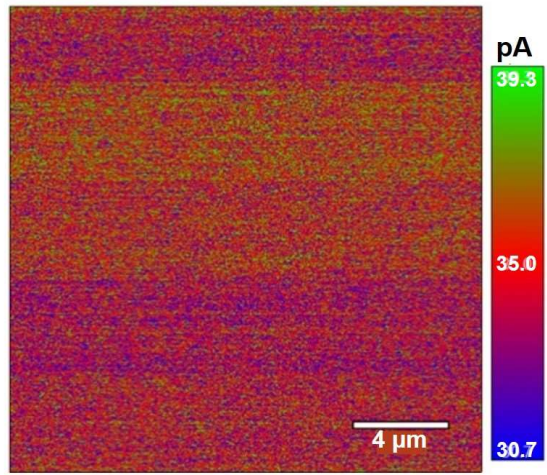

**Figure S3:** C-AFM scanning of barrier conductivity under an applied bias of +5V for 20nm thick PVDF layer.

## S4. Thermal strain effect

**Figure S4** displays the saturation magnetization ($M_s$) as a function of temperature for PVDF/Co composite and pure Co. We have observed an increase of magnetization about 10% from 10K to 300K for the Co on PVDF. This behavior is completely different from the normal decrease of $M_S$ with temperature for pure Co due to the magnetization disorder resulted by the Curie temperature. In fact, the increase of Co $M_s$ is related to the magnetostriction effect of Co [**Erreur ! Source du renvoi introuvable.**] because PVDF has an anisotropic thermal expansion, which can generate a uniaxial strain on Co when changing the temperature [**Erreur ! Source du renvoi introuvable.**]. The thermal strain can then effectively influence on the junction resistance, as already demonstrated in many reports [**Erreur ! Source du renvoi introuvable.**]. The thermal strain in the Co/Au electrode could also be related to the low temperature growth procedure for our OMFTJ, especially the deposition of Co/Au at low temperature.

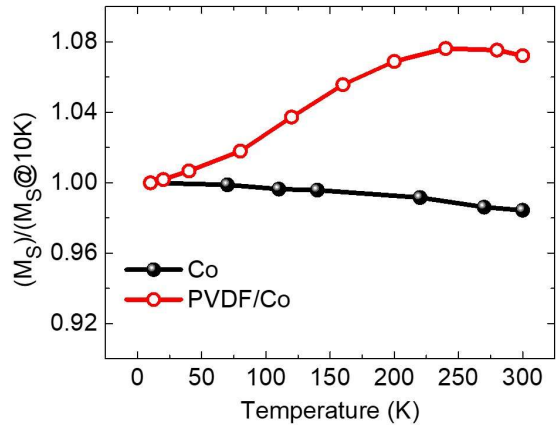

**Figure S4:** Saturation magnetization as a function of temperature for PVDF/Co composite and pure Co.

## S5. Temperature dependence of transport behavior with two polarization methods

We have measured the temperature dependence of resistance and magneto-transport behavior for a thin device of LSMO/PVDF(13nm)/Co with two polarizing setting methods: 1) states polarized only at 20K and 2) states polarized at each measuring temperature. In the first polarizing setting method, we have firstly ferroelectrically polarized PVDF in the negative polarization state at 20K, and then heated the sample to measure the magneto-response loops at different temperature. After reaching 300K, the sample is again cooled down to 20K and was polarized in the positive polarization state to perform temperature dependent measurement. In the second polarizing setting method, the PVDF is polarized at each measuring temperature in both states before measuring the magneto-response loops.

For both polarization methods, all MR curves decrease with the increase of temperature (**Figure S5(a)**). This validates that the diminishing of MR is mainly due to the loss of surface spin-polarization of LSMO electrodes but not due to the polarizing states of PVDF. For the temperature dependent resistance, as shown in **Figure S5(b)**, it is found that the difference between resistances in both polarization states decreases with temperature for the first method, but the difference is enlarged for the second method. This means that the second method can achieve better polarization states at higher temperature. Thermal activation of the fluctuation of ferroelectric domain should happen for the first method. The more evidence change can be observed in the negative polarization state, which means that positive polarization state is more stable with temperature. As a consequence, we have observed an increase of TER with increasing temperature for the second method (**Figure S5(c)**). The TER is much enhanced and reaches about 200% at RT with the second polarization method. This study gives a strong argument that the decreasing of TER for the temperature range higher than 120K is mainly due to the thermal fluctuation on the ferroelectric domains.

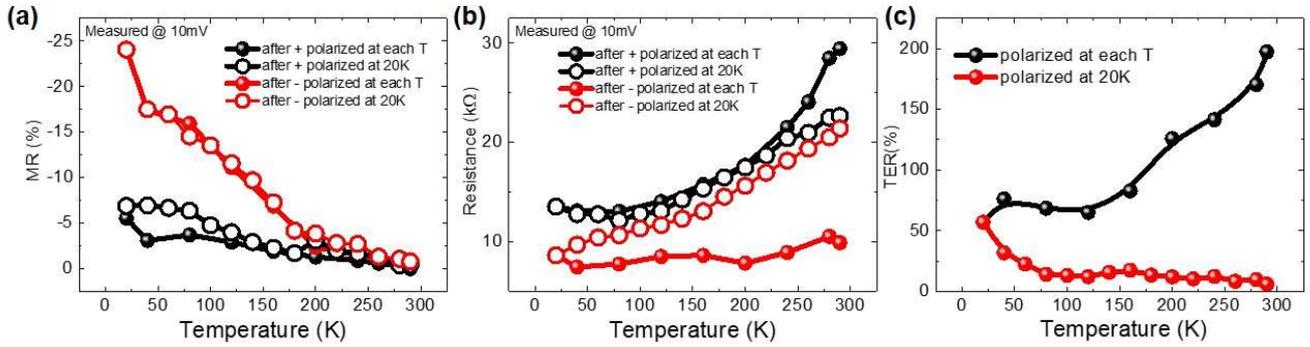

**Figure S5:** (a) Temperature dependence of MR after positive(black) and negative(red) polarization. The solid circles show the MR measured after being polarized at each temperature, the opened circles show those after being polarized only at 20K. (b) Temperature dependence of $R_P$ after positive(black) and negative(red) polarization. The solid circles show the $R_P$ measured after being polarized at each temperature, the opened circles show those after being polarized only at 20K. (c) Temperature dependence of TER after being polarized at each temperature (black) and at 20K (red).